\documentstyle[12pt]{article}

\def\balpha{{\mbox{\boldmath $\alpha$}}}
\def\bbeta{{\mbox{\boldmath $\beta$}}}
\def\bgamma{{\mbox{\boldmath $\gamma$}}}
\def\be{\begin{equation}}
\def\ee{\end{equation}}

\input epsf

\begin{document}
\rightline{CU-TP-821}
\rightline{hep-th/yymmddd}
\rightline{\today}
\vskip 1cm
\centerline{\large\bf BPS Monopoles and Electromagnetic
Duality\footnote{To appear in the Proceedings of the Inaugration
Conference of APCTP, Seoul, Korea (1996).}}
\vskip 0.7cm

\centerline{\it
Kimyeong Lee\footnote{electronic mail: klee@phys.columbia.edu} }

\vskip 3mm
\centerline{Physics Department, Columbia University, New York, NY, 10027}
\vskip 0.5cm
\centerline{\bf ABSTRACT}
\vskip 2mm
\begin{quote}
We review our recent work on the BPS magnetic monopoles and its
relation to the electromagnetic duality in the $N=4$ supersymmetric
Yang-Mills systems with an arbitrary gauge group. The gauge group can
be maximally or partially broken.  The low energy dynamics of the
massive and massless magnetic monopoles are approximated by the moduli
space metric.  We emphasize the possible connection between the nature
of the monopole moduli space with unbroken gauge group and the physics
of mesons and baryons in QCD.

\end{quote}

\setcounter{footnote}{0}

\section{Introduction}

Recently there has been a considerable progress in our understanding
of the electromagnetic duality. This duality is a duality between
strong and weaking interacting theories and so intrinsically
nonperturbative. Especially a simple case appears when we consider the
$N=4$ supersymmetric Yang-Mills systems which can be regarded
self-dual under the electromagnetic duality. (For the recent detailed
review, see Ref.\cite{olive}.) Thus, the spectrum of magnetic
monopoles should match to that of electrically charged particles
exactly.

In this talk, I will described some recent
works\cite{klee1,klee2,klee3} done in collaboration with Erick
Weinberg and Piljin Yi, concerning the low energy dynamics of magnetic
monopoles in the systems with larger gauge group than $SU(2)$. The
gauge symmetry can be maximally broken to abelian subgroups or
partially broken with unbroken nonabelian gauge group.

The outline of this talk is as follows: In Sec.2, we briefly review
the magnetic monopoles and the electromagnetic duality. In Sec.3, we
consider the BPS magnetic monopole configurations. We discuss zero
modes of the BPS configurations and show that the duality of the
electric and magnetic sectors entails the bettter understanding of the
low energy dynamics of the BPS monopoles.  In Sec.~4, we discuss how
the low energy dynamics of the BPS monopoles are approximated by the
moduli space dynamics, or the zero mode dynamics. I briefly describe
how to derive the moduli space metric for the distinct fundamental
monopoles.  In Sec.~5, two specific examples $SU(3)\rightarrow U(1)^2$
and $SU(4)\rightarrow U(1)^2\times SU(2)$ are studied. In Sec.~6, we
close with some concluding remarks.

\section{Duality}

The duality between electricity and magnetism has been fascinating us
quite a while. Initially, it has originated from the invariance of the
free Maxwell equations under the global phase rotation $({\bf E}+i
{\bf B})$ $\rightarrow$ $e^{i\alpha}$ $({\bf E} +i{\bf B})$. When the
electric and magnetic currents, $j_e^\mu$ and $j_m^\mu$, are
introduced, the Maxwell equation becomes
\begin{eqnarray}
& & \nabla \cdot ({\bf E}+i{\bf B}) = j_e^0 + i j_m^0,\\
& & \partial_t ({\bf E} + i{\bf B}) -i \nabla({\bf E}+ i {\bf B}) =
{\bf j}_e + i {\bf j}_m.
\end{eqnarray}
They are still invariant if we also rotate the four current $j^\mu_e +
i j_m^\mu$ $\rightarrow $ $e^{i\alpha}(j^\mu_e + i j_m^\mu)$. This
duality allows us to understand the classical interaction between
point particles carrying electric and magnetic charges.

Dirac\cite{dirac} showed that quantum mechanical interaction between
electrically charged particles and magnetic monopoles can be
consistently implemented only if the quantization law,
\begin{equation}
q g = 2\pi n \hbar
\end{equation}
with an integer $n$, between electric charge $q$ and magnetic charge
$g$ is satisfied. This quantization is also consistent with the
angular momentum quantization.  Not only this quantization law
explains the electric charge quantization in a presence of a single
monopole anywhere in the universe, it is also invariant under a more
restricted `electromagntic' duality, $(E,B)\rightarrow (B,-E)$ and
$(j_e^\mu,j_m^\mu)\rightarrow (j_m^\mu, -j_e^\mu)$. In terms of the
electric charge unit $e$, the minimum magnetic charge should be
$2\pi\hbar/e$. Thus when the electromagnetic coupling constant is
small, the magnetic coupling constant is large, and vice versa.

By considering two interacting dyons of charge $(q_i,g_i) $ with
$i=1,2$, Schwinger and Zwanziger\cite{schwinger} extended the Dirac's
law to
\begin{equation}
q_1g_2 -q_2g_1 = 2\pi \hbar.
\end{equation}
By considering the effect of the $CP$ violating $\theta$ term in the
Maxwell systems and its parent Yang-Mills-Higgs systems,
Witten\cite{witten1} showed that the above quantization law can be
implemented by pure electrically charged particles of charge in unit
of $e$ and and dyons of quantized magnetic charge $g=(2\pi/e)n_m$ and
fractional electric charge
\begin{equation}
q = e( n_e + \frac{\theta}{2\pi}n_m),
\end{equation}
where $n_e,n_m$ are integers.

On the other hand, t'Hooft and Polyakov\cite{thooft} found magnetic
monopoles as solitons in Yang-Mills Higgs systems where the gauge
group $SU(2)$ is broken to $U(1)$ by the Higgs mechanism. The magnetic
charge is topologically quantized $g=4\pi/e$ where $e$ is the charge
of elementary charged vector bosons.  When the potential of the Higgs
field vanishes, Bogomolny and others\cite{bogomolny} found a bound on
the energy by electric and magnetic charges $ (q,g) $,
\begin{equation}
E \ge v \sqrt{g^2+q^2},
\end{equation}
which is saturated by every elementary particle and dyons. In this
system, there exist also elementary massive charged vector bosons
$W_\mu^{\pm}$ whose mass saturate the above bound.

Montonen and Olive\cite{montonen} proposed that in this theory holds
the electromagnetic duality which transforms $(e,g)\rightarrow (g,-e)$
and $(n_e,n_m)\rightarrow (n_m,-n_e)$. Soon it was realized by
Osborn\cite{osborn} that to match the spectrum of the electric sector
to that of the magnetic sector in spin content, one needs the $N=4$
supersymmetry. In this case both electric and magnetic sectors
saturate the Bogomolny bound and so belong to the short representation
of the supersymmetry with the maximal spin one\cite{witten2}.

The $N=4$ supersymmetric theory is finite and there is no running
coupling constant and so the classical bound is expected to remain
exact quantum mechanically.  When the initial coupling constant is
small, elementary particles are interacting weakly and magnetic
monopoles are strongly interacting.  As the size $\sim 1/(ev)$ and
mass $\sim v/e$ of magnetic monopoles are much larger than those of
elementary particles and we can approach the monopole physics
semiclassically.

Instead if we make the coupling constant to be big, the theory becomes
a strongly interacting gauge theory. In this case elementary particles
are interacting strongly 
and solitons are interacting weakly. The mass of monopoles becomes
smaller than that of elementary particles and the size of monopoles
becomes smaller than the Compton length of magnetic monopoles. Thus,
we cannot treat magnetic monopoles as classical solitons and so seem
to be lost.

However if the electromagnetic duality holds, there would be a weakly
interacting dual gauge theory where magnetic monopoles appear as
elementary particles and $W$ bosons appear as solitons.  As the dual
theory itself is the $N=4$ supersymmetric Yang-Mills theory, the dual
theory is identical to the original theory with the coupling constant
$4\pi/e$.

Thus the dynamics of strongly interacting $W$ bosons in the theory
with coupling constant $4\pi/e$ with small $e$ is identical to that of
strongly interacting magnetic monopoles in the theory with coupling
constant $e$. Thus the  understanding of strongly interacting magnetic
monopoles implies an understanding of strongly interacting $W$ bosons.

There are two main questions to be addressed in this context. First
question is about the validity of the duality. We do not know any
rigorous dual transformation of the theory. However we can still test
the duality. If the duality holds, the spectrum of electrically
charged particles should match that of magnetic monopoles. It would be
interesting to find other nontrivial tests of the duality. Second
question is about the implications of the duality. As I argued before,
the understanding of the magnetic monopole dynamics implies that of
strongly interacting elementary particles.

The electromagnetic duality can be generalized into several
directions. When the $\theta$ parameter is introduced, the
electromagnetic duality can be generalized to the $S$-duality.  In
this context, Sen\cite{sen} provided the evidence for the $S-$duality,
by finding the bound states of two identical magnetic monopoles with
odd number of electric charge.  Another direction is to consider the
more general gauge group than $SU(2)$. With a larger gauge group one
can consider the case where the unbroken gauge group is not purely
abelian.  When the duality is generalized to the unbroken nonabelian
gauge group\cite{goddard1}, its meaning becomes more subtle as we will
see later.

\section{BPS Monopoles}

As emphasized before, the magnetic monopole configurations in the $N=4$
supersymmetric Yang-Mills theory are described by the BPS magnetic
monopoles. Since there are six  independent scalar fields in the
system, the magnetic monopole dynamics is rather different at the
generic points of the symmetry breaking. The monopole dynamics is rich
when the vacuum expectation values of the six scalar fields are
parallel and this is the case we will consider here. For more general
case, see a recent article\cite{hollowood}.

The energy of a general field configuration is bounded by the
topological quantity which is a function of its electric and magnetic
charge. The field configurations which saturate this BPS bound satisfy
the first order BPS equation.

We start with an arbitrary gauge group $G$ of rank $r$. Its generators
are made of $r$ commuting operators $H^s$ and raising and lowering
operators $E_{\pm\balpha}$.  We choose the vacuum expectation
value of the Higgs field as $\Phi_0 = {\bf h \cdot H}$ in the Carten
subalgebra. When the gauge symmetry is maximally broken to $U(1)^r$,
there is a unique set of simple roots $\bbeta_a$ such that $\bbeta_a
\cdot {\bf h}>0$. When the gauge symmetry is partially broken to
$K\times U(1)^{r-k}$ with a semisimple group $K$ of rank $k$, there
exists a unique set of simple roots ${\bbeta_a, \bgamma_i}$ modulo the
Weyl group of $K$, where $\bgamma_i$ are the simple roots of $K$ and
so  $\bgamma_i\cdot{\bf h} = 0$.

In the direction chosen to define $\Phi_0$, the asymptotic magnetic
field of $BPS$ configuration will be of form
\begin{equation}
    \vec{B} = {{\hat r} \over 4\pi r^2} {\bf g\cdot H},
\end{equation}
whose magnetic charge ${\bf g}$ satisfy the topological quantization
condition \cite{goddard2}
\begin{equation}
    {\bf g} = {4\pi \over e} \left[\sum_{a=1}^{r-k} n_a \bbeta_a^* +
\sum_{j=1}^k 
q_j \bgamma_j^* \right],
\label{gcoeff}
\end{equation} 
where $\balpha^* = \balpha/\balpha^2$ is the dual of the root
$\balpha$ and the $n_a$ and $q_j$ are non-negative integers. In the
gauge group $SU(N)$, we choose the normalization $\alpha^* =
\alpha$. The $n_a$'s are the topologically conserved charges.  For a
given solution they are uniquely determined and gauge invariant, even
though the corresponding $\bgamma_a$ may not be.  The $q_j$ are
neither gauge invariant nor conserved except their sum.

For maximal symmetry breaking to $U(1)^r$, there is a unique
fundamental monopole solution associated with each of the $r$
topological charges.  To obtain these, we first note that any root
${\mbox{\boldmath $\alpha$}}$ defines an $SU(2)$ subgroup generated by
the raising and lowering operators ${\bf E}_{\pm \balpha}$. One can
embed the $SU(2)$ magnetic monopole of unit charge to get a
spherically symmetric monopole solution for given root $\balpha$. For
each simple root $\bbeta_a$, the monopole has the unit topological
magnetic charge ${\bf g} = (4\pi/e)\bbeta_a^*$ and mass
$m_a=(4\pi/e){\bf h}\cdot \bbeta_a^*$.  All other BPS solutions can be
understood as multimonopole solutions containing $N= \sum_{a=1}^r n_a$
fundamental monopoles\cite{weinberg1}. Especially the rotationally
invariant monopole solutions for the composite positive roots
$\balpha$ is not fundamental.  
The moduli space for these multimonopole solutions has $4N$
dimensions, corresponding to three position variables and a single
$U(1)$ phase for each of the component fundamental
monopoles\cite{weinberg1}.

Matters are somewhat more complicated when the unbroken gauge group is
nonabelian \cite{weinberg2}.  If the long-range magnetic field has a
nonabelian component (i.e., if ${\bf g}\cdot {\mbox{\boldmath
$\gamma$}}_j)\ne 0$, the index theory methods used to count zero modes
in Refs.~\cite{weinberg1} and \cite{weinberg2} fail for technical reasons
related to the slow falloff of the nonabelian field at large
distance. If the total magnetic charge does not have the nonabelian
component so that ${\bf g}\cdot {\mbox{\boldmath $\gamma$}}_j =0$,
these difficulties do not arise and the number of normalizable zero
modes is
\begin{equation}
    N = 4 \left[ \sum_a n_a + \sum_j q_j \right] .
\end{equation}

Let us illustrate the above ideas in the case where the $SU(3)$ gauge
group is broken to either $U(1)^2$ or $SU(2)\times U(1)$. In the
maximally broken case, $\Phi_0={\bf h}\cdot {\bf H}$ and two simple
positive roots are $\bbeta,\bgamma$ as shown in Fig.~1. There are
fundamental monopoles corresponding to simple dual roots $\bbeta^*$
and $\bgamma^*$, but there is only a composite monopole for root
$\balpha$.  As ${\bf h}\rightarrow {\bf h}'$, the unbroken gauge group
becomes $SU(2)\times U(1)$ as ${\bf h}'\cdot \bgamma=0$. The unbroken
$SU(2)$ is associated with the raising and lowerling operators ${\bf
E}_{\pm \bgamma}$.  The minimum magnetic charge without the nonabelian
component is
\begin{equation} 
{\bf g} = \frac{4\pi}{e}(2\bbeta^*+ \bgamma^*)
\end{equation} 
because ${\bf g}\cdot \bgamma=0$, which are composed of two massive
and one massless monopoles. The number of zero modes for the above
magnetic charge is  $12$.

\begin{center}
\leavevmode
\epsfysize=3.0in\epsfbox{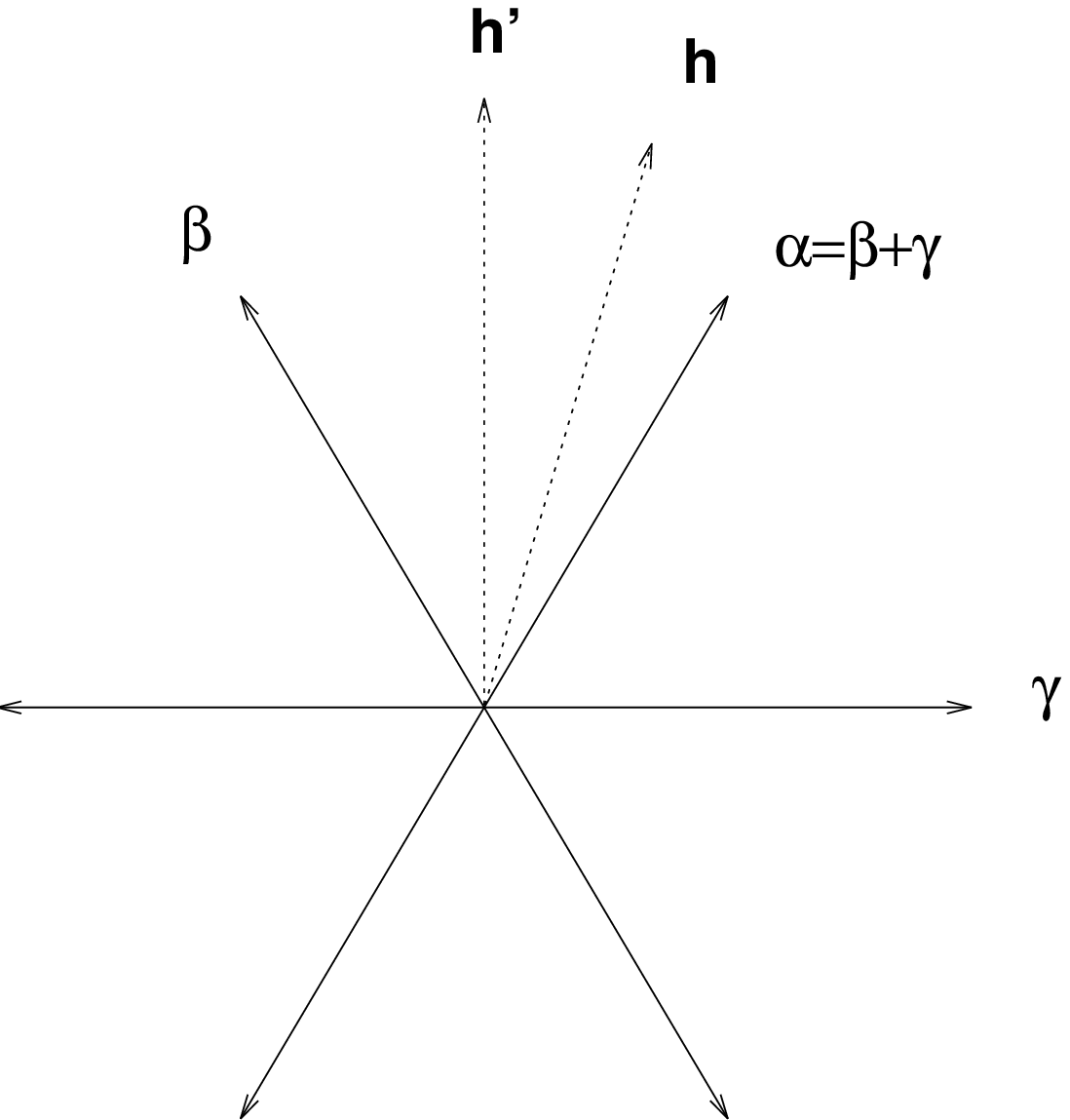}
\end{center}
\vskip 1cm
\begin{quote}
{\bf Figure 1:} {\small
Root diagram for $SU(3)$. When $\Phi_0={\bf h}\cdot {\bf H}$, the
symmetry is maximally broken. When $\Phi_0={\bf h}'\cdot {\bf H}$, the
symmetry is partially broken.}
\end{quote}
\vskip 0.6cm

In the maximally broken symmetry phase there are as many charged
vector bosons as the number of positive roots, whose number is much
larger than that of simple roots if the gauge group is bigger than
$SU(2)$.  On the other hand we just argued that classically the number
of fundamental monopoles is identical to that of simple roots. In the
above $SU(3)$ example there are fundamental monopoles corresponding to
the simple roots $\bbeta^*,\bgamma^*$, but not for the composite root
$\balpha^*$.

This seems to contradict with the duality hypothesis that the spectrum
of charged particles should match exactly that of magnetic monopoles.
However the duality is an intrinsically quantum mechanical statement:
one has to study quantum mechanical spectrum of magnetic monopoles,
which raises a possible existence of the quantum mechanical bound
state of fundamental monopoles for each composite root.  As the BPS
mass formula is expected to be exact even in quantum mechanically in
the $N=4$ supersymmetric theory, the bound energy of these composite
monopoles would be zero. It seems hard to find such threshold bound
states of magnetic monopoles directly in the full quantum field
theory. However, one can approximate the very low energy dynamics of
magnetic monopoles by the nonrelativistic moduli space dynamics, whose
quantum mechanics may allow such expected bound states. Hence, let us
now turn our attention to the magnetic monopole moduli space.

\section{Moduli Space Approximation}

The BPS configurations for a given magnetic charge, and so the same
energy, are parameterized by the $N$ collective coordinates $z_\alpha,
\alpha=1...N$, or moduli up to local gauge transformations.  The space
of gauge-inequivalent BPS configurations for a given magnetic charge,
\begin{equation}
A_\mu({\bf x};z_\alpha) = ({\bf A}({\bf x};z_\alpha),
A_4=\Phi({\bf x},z_\alpha)), 
\end{equation}
is called the moduli space of solutions. The  $N$
zero modes $\delta_\alpha A_\mu = \partial A_\mu /\partial
z_\alpha - D_\mu \epsilon_\alpha $ satisfy the linearized BPS equation
and are  chosen to satisfy the
background gauge 
\begin{equation}
D_\mu \delta_\alpha A_\mu=0
\label{back}
\end{equation}
with $\partial_4=0$.  When we consider the fluctuations around the BPS
magnetic monopole configurations, there are massless modes and massive
modes. If the initial energy is arbitrary small, the dynamics of BPS
monopoles may be approximated by that of moduli\cite{manton1}. The
initial field configuration at a given time will be characterized by
$A_\mu({\bf x},z_\alpha(t))$ and its time derivative, $\dot{z}_\alpha
\delta_\alpha A_\mu$ in the $A_0=0$ gauge. The Gauss law constraint on
the initial configuration is exactly the background gauge
(\ref{back}).

Since there is no force between monopoles at rest, one expect the low
energy dynamics is given by the kinetic part of the Yang-Mills-Higgs
Lagrangian. In the $A_0=0$ gauge, this becomes
\begin{equation}
L = \frac{1}{2} G_{\alpha\beta}(z_\alpha)  \dot{z}_\alpha
\dot{z}_\beta
\label{modu}
\end{equation}
where  $G_{\alpha\beta}(z_\alpha) =\int d^3x \,{\rm tr}\,
\delta_\alpha A_\mu \delta_\beta A_\mu $. While one
can study some characteristics of this metric, it is hard to obtain
directly from the BPS field configurations which themselves are not
known in general.  However some formal characteristics of the metric
can be deduced from this.  The important property of the metric is
that it is hyperk\"ahler. This property is also related to the field
theoretical supersymmetry which should be incorporated into the
Lagrangian (\ref{modu}) to be consistent \cite{blum}.

 The full moduli space and its metric are known for two identical
monopoles\cite{atiyah,klee1,connell,gauntlett}.  For $N>2$ the metric
for the case where all the component fundamental monopoles are all
distinct was given in Ref.~\cite{klee2}; for all other cases the
explicit form of the metric is known only for the region of moduli
space corresponding to widely separated fundamental
monopoles\cite{gibbons1,klee2}.

There are several approaches to calculate the moduli space metric.
Here we focus on the approach taken by Manton and
Gibbons\cite{manton2,gibbons1}. The metric determines the interaction
between BPS magnetic monopoles of low kinetic energy and vice
versa. Once we understand the interaction between magnetic monopoles,
we can deduce the metric. The interaction between $n$ fundamental BPS
magnetic monopoles becomes particularly simple when their mutual
distances are very large. In this large separation, the electric
charge of each monopole is conserved as the possible violating term is
exponentially small. Thus, it is easier to consider the interaction
between $n$ fundamental dyons in large separation. The interaction
between the dyons in large separation becomes purely electromagnetic
and scalar in nature.  Once the nonrelativistic Lagrangian for dyons
is obtained, the Lagrangian for monopoles can be obtained by the
Legendre transformation of the electric charge to the phase variables.

Specifically we consider the $r$ distinct fundamental monopoles
in the maximally broken gauge group $G$.  The $a$-th monopole
associated with the simple root $\bbeta_a$ has the position ${\bf x}_a$
and the phase variable $\xi_a$. The metric obtained from the method
mentioned previously is
\begin{equation}
{\cal G}=\frac{1}{2}M_{ab}d{\bf x}_a\cdot d{\bf x}_b+\frac{g^4}{2(4\pi)^2} 
(M^{-1})_{ab}(d\xi_a+{\bf W}_{ac}\cdot d{\bf x}_c)(d\xi_b+{\bf W}_{bd}
\cdot d{\bf x}_d) ,\label{metric}
\end{equation}
where
\begin{eqnarray}
M_{aa} &=& m_a  - \sum_{c\ne a} \frac{g^2 \bbeta_a^* 
\cdot \bbeta_c^*}{ 4\pi r_{ac}},\nonumber \\
M_{ab} &=&\frac{g^2 \bbeta_a^* \cdot 
\bbeta_b^*}{ 4\pi r_{ab}}\qquad
\hbox{\hskip 1cm if $a\neq b$},
\end{eqnarray}
with $m_a=g\,\bbeta_a^*\cdot {\bf h}$, and
\begin{eqnarray}
{\bf W}_{aa}&=&-\sum_{c\neq a}\bbeta_a^*\cdot
\bbeta_c^*{\bf w}_{ac},\nonumber\\
{\bf W}_{ab}&=&\bbeta_a^*\cdot
\bbeta_b^*{\bf w}_{ab}\qquad
\hbox{\hskip 1cm if $a\neq b$}.
\end{eqnarray}
with ${\bf w}_{ab}={\bf w}({\bf x}_a-{\bf x}_b)$ being value at ${\bf
x}_a$ of the Dirac potential due to the $b$-th monopole. The
$q$-independent part of the monopole rest energies has been omitted.
The fact that this asymptotic metric is hyperk\"ahler can be shown
trivially, following the argument by Gibbons and
Manton\cite{gibbons1}. The key ingredient is that $\nabla 1/r = \nabla
\times {\bf w}(r)$. The question is whether the asymptotic metric when
it is extended in the interior region is nonsingular.

Neither of these objections arises for the moduli space corresponding
to a collection of several distinct fundamental monopoles in a larger
group, provided that each corresponds to a different simple root. In
this case, one can show that the metric is nonsingular everywhere by
going to the center of mass frame\cite{klee2}. The metric has the
right isometry: the rotational symmetry and the $U(1)$ symmetry for
each conserved $U(1)$ charge. More recently it has been strongly
argued that the metric in this case is indeed
exact\cite{murray}.

\section{Examples of the Moduli Space Metric}

Here we discuss in detail two simple examples whose moduli space is
known. First we discuss the case where the gauge symmetry is maximally
broken to the abelian subgroups. Then we discuss the case where the
symmetry is partially broken with unbroken nonabelian group.

\subsection{$SU(3)\rightarrow U(1)^2$}

For two distinct fundamental monopoles $M_{\bbeta} $ and $M_{\bgamma}
$ with the Dynkin diagram shown in Figure 1, the moduli space metric
has been obtained from Eq.(\ref{metric}). In terms of the center of
mass coordinates ${\bf R},\chi$ and the relative coordinates ${\bf
r},\psi$, the geometry of the center of mass coordinates is shown to
be $R^4$ and that of the relative
coordinate\cite{connell,gauntlett,klee1} is the Taub-NUT space with
the metric
\begin{equation}
{\cal G}_{\rm rel}^{(2)} = \mu\left( (1+ \frac{2l}{r})\,d{{\bf r}}^2 
+ 4l^2(1+ \frac{2l}{r})^{-1}(d{\psi}+ {\bf w}({\bf r})\cdot d{{\bf
r}})^2 \right). 
\label{gtwo}
\end{equation}
The Taub-NUT metric is smooth everywhere including the origin if the
$\psi$ has a period of $4\pi$, which is true as one can see from the
charge quantization.

In addition, there are identification maps forming the integer group
$Z$ on the cylinder $(\chi,\psi)$. The result is that the total moduli
space is of the form \cite{klee1}
\begin{equation}
{\cal M } = R^3 \times \frac{R^1 \times {\cal M}_0}{Z} ,  \label{Roneform}
\end{equation}
where ${\cal M}_0$ is the Taub-NUT manifold.

The relative metric possesses the $SU(2)$ rotational symmetry and the
global $U(1)$ symmetry, which are required from the physics of two
distinct dyons. Our metric is obtained from the interaction between
monopoles in large separation.  However there is a complete
classification of four dimensional hyperk\"ahler spaces in four
dimensions with the rotational symmetry acting on three dimensional
space\cite{atiyah}. Among them only one whose symmetry and asymptotic
form match with those of our our relative moduli space is the Taub-NUT
space itself.  Thus the asymptotic form of the metric for the distinct
monopoles turns out to be exact everywhere.

We now have the moduli space approximation of the low energy dynamics
of two monopoles $M_{\bbeta^*}$ and $M_{\bgamma^*}$, and so let us
come back to the electromagnetic duality in this theory, which was
discussed in Sec.3. First of all we should supersymmetrize our
effective action to describe the $M_{\bbeta}$ and $M_{\bgamma}$ in the
$N=4$ supersymmetric theory\cite{blum}. The ground state of such a
supersymmetric Hamiltonian is given by the normalizable self-dual
two-form. In the Taub-NUT space there exists unique such a two-form,
$\Omega = dQ,$ where $Q= (1+2l/r)^{-1}(dq+{\bf w}({\bf r})\cdot d{\bf
r})$ is the one-form associated with the conserved relative
charge\cite{gauntlett, klee1}.
This ground state is interpreted naturally as the threshold bound
state corresponding to the fundamental monopole $M_{\balpha}$, whose
existence is necessary for the electromagnetic duality to hold.

The moduli space of distince monopoles in larger group has been
discussed in detail in Ref.[2]. The threshold bound state of these
monopoles has been found by Gibbons\cite{gibbons3}.

\subsection{$SU(4)\rightarrow U(1)^2\times SU(2)$}

Let us now consider the case where the $SU(4)$ gauge symmetry is
partially broken to $U(1)^2\times SU(2)$. In the maximally broken
case, the simple roots $\balpha,\bbeta,\bgamma$ are chosen so that
their inner products with ${\bf h}$ are positive. The root diagram of
$SU(4)$ is shown in Fig.2.  We take the limit where ${\bf h}\cdot
\bbeta =0$ so that the unbroken $SU(2)$ symmetry is associated to the
$\bbeta$ root.

\begin{center}
\leavevmode
\epsfysize=0.7in\epsfbox{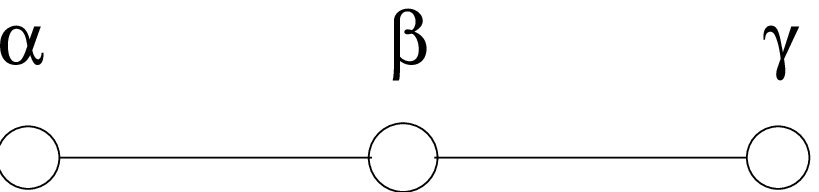}
\end{center}
\vskip 1cm
\begin{quote}
{\bf Figure 2:} 
{\small
Root diagram for $SU(4)$.} 
\end{quote}
\vskip 0.6cm

Let us consider the duality in this case.  The
supersymmetric multiplets of elementary charged and neutral particles
can be listed in a four-by-four hermitian matrix as follows:
\vskip 0.3cm 
\be 
\left( \begin{array}{ccc} \gamma & W_\balpha &
W_{\balpha+\bgamma} \\ 
W^*_\balpha & {\rm gluons} & W_\bgamma \\
W^*_{\balpha + \bgamma}  & W^*_\bgamma & \gamma' \\
\end{array} \right).
\ee
\vskip0.2cm
The diagonal elements are made of two photons and a gluon of the
unbroken $SU(2)$ gauge group. The vector boson  $W_\bbeta$ is a part
of the gluon spectrum for the unbroken $SU(2)$ gauge group.

The charged vector bosons $W_\balpha$ and $W_\bgamma$ belong to the
spinor representation of $SU(2)$. There are  also  charged vector bosons
$W_{\balpha+\bgamma}$, which are  neutral under the $SU(2)$ gauge group. In
the magnetic sector, there are two massive fundamental monopoles
$M_\balpha$ and $M_\bgamma$. Also there is a massless monopole
$M_\bbeta$, which can be regarded as the dual of the massless gluon
$W_\bbeta$. The magnetic charge sector are shown as follows:
\be
\left(\begin{array}{ccc} 
     \gamma & M_\balpha & ?  \\
    M^*_\balpha  & {\rm gluons} & M_\bgamma \\
    ?  & M^*_\bgamma & \gamma'
\end{array} \right). 
\ee

The classical monopole configuration corresponding to  the question
mark  has the magnetic charge  
\be
{\bf g} = \frac{4\pi}{e} (\balpha+ \bbeta + \bgamma),
\label{nonabelian} 
\ee
which does not have any nonabelian component of magnetic charge. While
this configuration has the right quantum number as the dual
configuration for $W_{\balpha+\bgamma}$, which is color neutral, it is
a composite of two massive and one massless monopoles and so has 12
zero modes. For the duality to holds even in the case with unbroken
nonabelian symmetry, there should be a unique state in the quantum
mechanics of the above monopole configuration, which may be realized
as a threshold bound.

Similar to the $SU(3)$ case, we want to see this bound state, if
exists, in the low energy effective Lagrangian, which is described by
the moduli space metric. The metric of the moduli space in this case
is obtained by taking the massless limit of the metric
(\ref{metric}). After separating out the flat center-of-mass metric,
the metric of the relative moduli space is $8$ dimensional.  The
metric is written in terms of the relative position vector ${\bf r}_1$
between the $\balpha-\bbeta$ monopoles, the ${\bf r}_2$ between the
$\bbeta-\bgamma$ monopoles, and the two relative phases
$\psi_1,\psi_2$. The resulting metric is the so-called Calabi-Taubian
metric\cite{klee3, gibbons2},
\begin{equation}
ds^2 = C_{AB} d{\bf r}_A\cdot d{\bf r}_B + C_{AB}^{-1} ( d\psi_A +
{\bf w}({\bf r}_A) \cdot d{\bf r}_A) ( d\psi_B +
{\bf w}({\bf r}_B) \cdot d{\bf r}_B)
\end{equation}
where 
\begin{equation}
C_{AB} = \left( \begin{array}{cc}
                 \mu + 1/r_1 & \mu \\
                 \mu  & \mu + 1/r_2  
                 \end{array} \right)
\end{equation}                          
where $\mu$ is related to the reduced mass of two massive monopoles
and the coupling $e$ is chosen so that the metric appears simple.

The isometry of the moduli space consists of the rotational group and
the unbroken gauge group, under which the monopole kinetic energy is
invariant.  The isometry of the relative metric is then made of
$U(1)\times SU(2)_{gauge} \times SU(2)_{rot}$.  The $8$ relative
coordinates change their meaning in the massless limit: \hfill\break
(1) Three of them are ${\bf r}_1 + {\bf r}_2$, which describes the
$SU(2)_{gauge}$ gauge invariant relative position between to massive
monopoles $\balpha$ and $\bgamma$. \hfill\break (2) One of them is the
conjugate phase of the $SU(2)_{gauge}$ invariant relative charge $q_1
+ q_2$ of two massive monopoles. \hfill\break (3) One of them is $a =
r_1 + r_2 $ which is the $SU(2)_{gauge}$ invariant length parameter
and is the total distance of line connecting $\balpha,\bbeta,\bgamma$
monopoles \hfill\break (4) Three of them make the three dimensional
gauge orbit of $SU(2)$.

The structure of the monopole configuration can be learned by studying
the $SO(5)$ case, where the explicit monopole solution is
known\cite{weinberg3}. The parameter $a$ characterizes the size of the
massless nonabelian cloud around the massive monopoles. In our case,
the profile of the size is given by the ellipsoid, whose focal points
are two massive monopoles.  Figure 3 shows such a three magnetic monopole
configuration. The nonabelian magnetic charge of two massive monopoles
got shielded outside the ellipsoid parameterized by $a$.

\begin{center}
\leavevmode
\epsfysize=2.0in\epsfbox{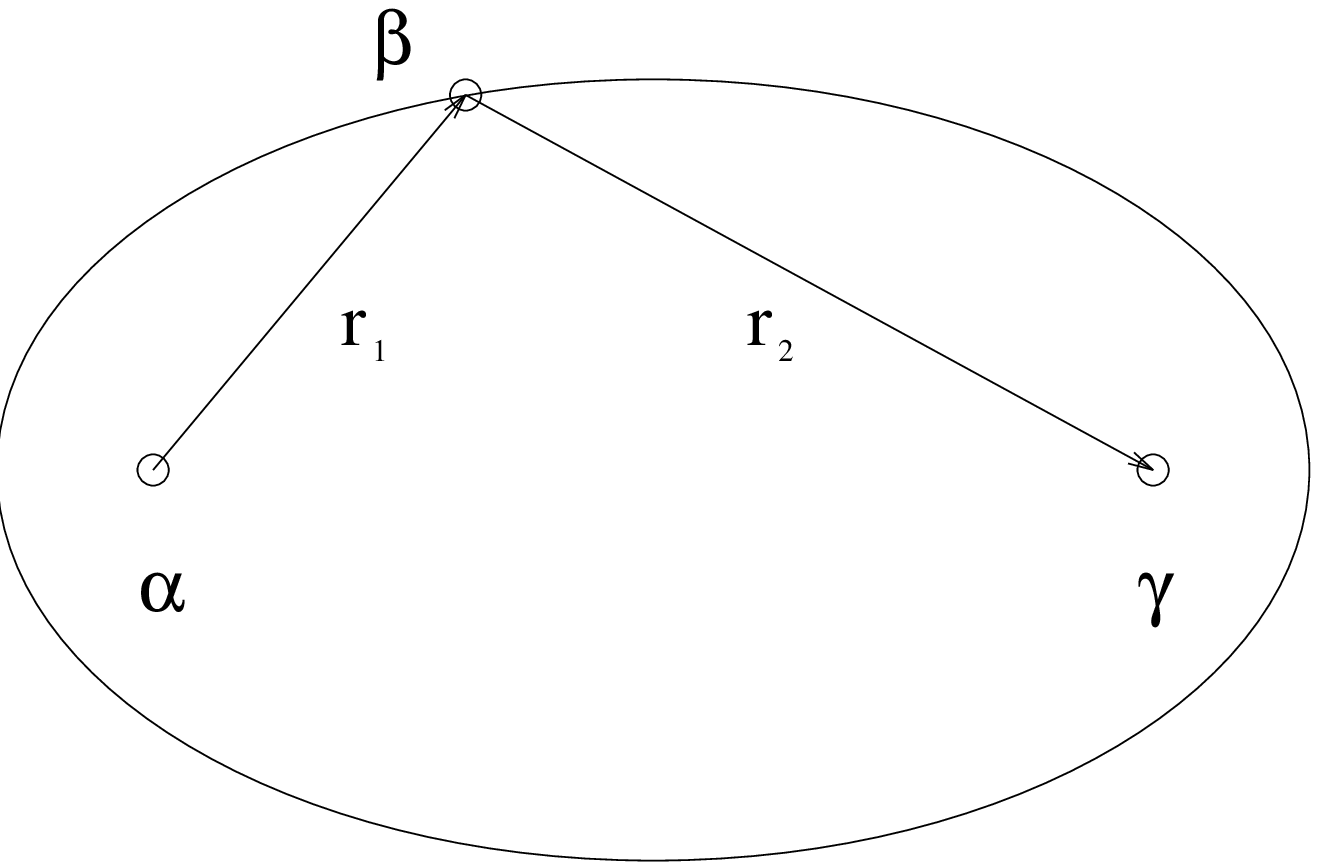}
\end{center}
\vskip 1cm
\begin{quote}
{\bf Figure 3:} {\small The $\balpha+\bbeta+\bgamma$ system in
the case $SU(4)\rightarrow U(1)^2\times SU(2)$.} 
\end{quote}
\vskip 0.6cm

Once the moduli space is known, we ask whether there exists a unique
threshold bound state of the magnetic charge(\cite{klee3}).  Once such
bound is found, the electromagnetic duality can be said to hold even
when the unbroken symmetry is partially nonabelian. However, there is
a reason to doubt its existence. When two massive monopoles overlap
each other, the relative moduli space is $R^4$, which does not have
any the normalizable self-dual two-form. If there exists no threshold
bound state, the nature of the duality has to be reexamined more
carefully.

The moduli space teaches us something else too. The massive magnetic
monopoles, $M_{\balpha}, M_{\bgamma}$ belong to the fundamental
representation of the magnetic $SU(2)$ group. We need to add some
massless monopoles, which belongs to the adjoint representation of the
unbroken simple group, to make the configuration to be a magnetic
gauge singlet. The moduli space metric, more precisely, the classical
monopole field configuration, tells us how massless monopoles behave
around the massive monopoles. While there is no magnetic flux
confinement, the massless monopole configuration looks like that of a
string connecting two massive monopoles. This is a dual version of
mesons, where quarks are replaced by massive monopoles and gluons are
replaced by massless monopoles.

The above consideration can be extended further.  Let us consider the
example where $SU(4)\rightarrow SU(3)\times U(1)$ so that
$M_{\balpha}$ is massive and $M_{\bbeta},M_{\bgamma}$ are massless
with the Dynkin diagram shown in Figure 2. The magnetic charge without
the nonabelian magnetic component is
\begin{equation}
{\bf g} = \frac{4\pi}{e}(3\balpha+2\bbeta+\bgamma).
\end{equation}
While we do not know the moduli space in this case, we can deduce a
few facts about this configuration. The number of zero modes
corresponding to massless monopoles is 12, among which 8 will be the
dimension of the gauge orbit of $SU(3)$. The rest 4 will be the gauge
invariant cloud shape parameters.  The massive magnetic monopoles
belong to the triplet of the unbroken $SU(3)$ magnetic group. Thus
this configuration is that of a proton in the dual picture, where
again quarks correspond to massive monopoles and gluons to massless
monopoles. If we understand how the massless monopoles are arranged,
we may have a better understanding of the confining string profile
inside a proton, assuming that the mass of quarks is much larger than
the QCD scale and their mutual distance is larger than that of
confinement.

In the case where $SU(3) \rightarrow SU(2)\times U(1)$ with magnetic
charge in Eq.(10), the magnetic moduli space has been found somewhat
earlier by Dancer\cite{dancer}. This moduli space of this
configuration of two identical massive monopoles and one massless
monopole is  somewhat similar to the $SU(4)\rightarrow SU(2)\times
U(1)^2$ case. While the metric is more complicated when two massive
monopoles are close to each other, we expect that  the nonabelian
cloud may  be arranged into an ellipsoid shape when two massive
monopoles get separated in large distance. This would be a dual
version of the baryon in the theory with $SU(2)$ gauge group. In the
dual picture the two identical massive monopoles would the identical
quarks in the spinor representation. It may be worthy to explore
further this moduli space in our context.

\section{Concluding Remarks and Discussions}

In this talk we have reviewed our recent work on the electromagnetic
duality in the $N=4$ supersymmetric Yang-Mills theories. We have
considered more general gauge group of higher rank, which is broken
spontaneously to maximally or partially. To match the magnetic
monopoles spectrum to that of electric charges, we need to
understand the low energy dynamics of BPS magnetic monopoles by the
moduli space approximation. Some magnetic monopole states, which are
needed for the duality to hold, appear only quantum mechanically as
threshold bounds of classical fundamental monopoles. We discuss in
detail the $SU(3)\rightarrow U(1)^2$ and $SU(4)\rightarrow
U(1)^2\times SU(2)$ cases.

There are some closely related ideas I have not discussed here for the
lack of time and space, which can be seen in the original papers.
Most interestingly there are a lot detailed discussion of the case
where the unbroken group has a nonabelian component. We believe the
further investigation along this direction would be fruitful.

\vskip 1cm
\centerline{\large\bf Acknowledgments}
\vskip  5mm

I like to thank the organizers of this conference. This work is
supported in part by the NSF Presidential Young Investigator program.

\end{document}